\documentstyle{article}
\include{graphicx}
\include{epsfig}

\begin{document}

\title{Apparent discordant redshift QSO-galaxy associations}

\author{Mart\'\i n L\'opez-Corredoira\\
Instituto de Astrof\'\i sica de Canarias\\ 
C/.V\'\i a L\'actea, s/n\\
ES-38200 La Laguna, Tenerife (Spain)\\
E-mail: martinlc@iac.es}

\maketitle

{\bf \large ABSTRACT}

An ``exotic'' idea proposed by Viktor Ambartsumian was that new galaxies 
are formed through the ejection from older active galaxies. Galaxies 
beget galaxies, instead of the standard scenario in which galaxies stem 
from the evolution of the seeds derived from fluctuations in the initial 
density field. This idea is in some way contained in the speculative proposal 
that some or all QSOs might be objects ejected by nearby galaxies, and that 
their redshift is not cosmological (Arp, G./M. Burbidge and others).

I will discuss some of the arguments for and against this scenario; in 
particular, I shall talk about the existence of real physical connections 
in apparently discordant QSO-galaxy redshift associations. On the one hand, 
there are many statistical correlations of high-redshift QSOs and nearby 
galaxies that cannot yet be explained in terms of gravitational lensing, 
biases, or selection effects; and some particular configurations have very 
low probabilities of being a projection of background objects. Our 
understanding of QSOs in general is also far from complete. On the other 
hand, some cases which were claimed to be anomalous in the past have found 
an explanation in standard terms. As an example, I will show some cases of 
our own research into this type: statistics of ULXs around nearby galaxies, 
and the Flesch \& Hardcastle candidate QSOs catalog analysis. My own conclusion 
is neutral.

\

\section{The problem and the observations that give rise to it}

Viktor A. Ambartsumian suggested the idea that new galaxies are formed 
through ejection from older active galaxies (Ambartsumian 1958). This idea 
has had a certain continuity in the research carried out over the last 
40 years based on the hypothesis that some extragalactic objects, and in 
particular high redshift QSOs, might be associated with low redshift galaxies, 
thus providing a non-cosmological explanation for the redshift in QSOs (e.g., 
Arp 1987, 2003; Narlikar 1989; Burbidge 2001; Bell 2002a,b, 2006, 2007; 
L\'opez-Corredoira \& Guti\'errez 2006a); that is, a redshift produced by a 
mechanism different from the expansion of the Universe or the Doppler effect. 
Ambarsumian never accepted the idea of non-cosmological redshifts; however, 
the scenario of QSOs ejected by galaxies is a common theme of the Armenian 
astrophysicist and in proposals of discordant QSO-galaxy redshift associations.

There are plenty of statistical analyses (e.g., Chu et al. 1984; Zhu 
\& Chu 1995; Burbidge et al. 1985; Burbidge 1996, 2001; Harutyunian 
\& Nikogossian 2000; Ben\'\i tez et al. 2001; Gazta\~naga 2003; 
Nollenberg \& Williams 2005; Bukhmastova 2007) showing an excess of 
high redshift sources near low redshift galaxies, positive and very 
significant cross-correlations between surveys of galaxies and QSOs, an 
excess of pairs of QSOs with very different redshifts, etc. An excess of 
QSOs near the minor axes of nearby parent galaxies has also been observed 
(L\'opez-Corredoira \& Guti\'errez 2007); however, the discovered excess for 
position angles lower than 45 degrees is significant only at the 3.5-$\sigma $ 
level (3.9-$\sigma $ for $z_{QSO}>0.5$) with the QSOs of the SDSS-3rd release 
(L\'opez-Corredoira \& Guti\'errez 2007) and somewhat lower [2.2-$\sigma $ 
(2.5-$\sigma $ for $z_{QSO}>0.5$)] with the SDSS-5th release.

There are plenty of individual cases of galaxies with an excess of QSOs with 
high redshifts near the center of nearby galaxies, mostly AGN. In some cases, 
the QSOs are only a few arcseconds away from the center of the galaxies. 
Examples are NGC 613, NGC 1068, NGC 1097, NGC 3079, NGC 3842, NGC 6212, NGC 
7319 (separation galaxy/QSO: 8"), 2237+0305 (separation galaxy/QSO 0.3"), 
3C 343.1 (separation galaxy/QSO: 0.25"), NEQ3 (see Fig. 1/left; a QSO-``narrow 
emission line galaxy'' pair separated 2.8" from another emission line galaxy 
with a second redshift, and all of them lying along the minor axis of an 
apparently distorted lenticular galaxy at $\sim $17" with a third redshift), 
etc. In some cases there are even filaments/bridges/arms apparently connecting 
objects with different redshift: in NGC 4319+Mrk 205, Mrk273, QSO1327-206, 
NGC 3067+3C232 (in the radio), NGC 622, NGC 3628 (in X-ray and radio), 
NEQ3 (Fig. 1/left), etc. The probability of chance projections of 
background/foreground objects within a short distance of a galaxy or onto 
the filament is as low as 10$^{-8}$, or even lower. 
The alignment of sources with different redshifts also suggests that they 
may have a common origin, and that the direction of alignment is the direction 
of ejection. This happens with some configurations of QSOs around 1130+106, 
3C212, NGC 4258, NGC 2639, NGC 4235, NGC 5985, GC 0248+430 (Fig. 1/right), etc. Other proofs presented in favor of the QSO/galaxies association with different redshift is that no absorption lines were found in QSOs corresponding to foreground galaxies (e.g. PKS 0454+036, PHL 1226), or distortions in the morphology of isolated galaxies.

\begin{figure*}
\vspace{1cm}
{\par\centering \resizebox*{4.9cm}{6cm}{\includegraphics{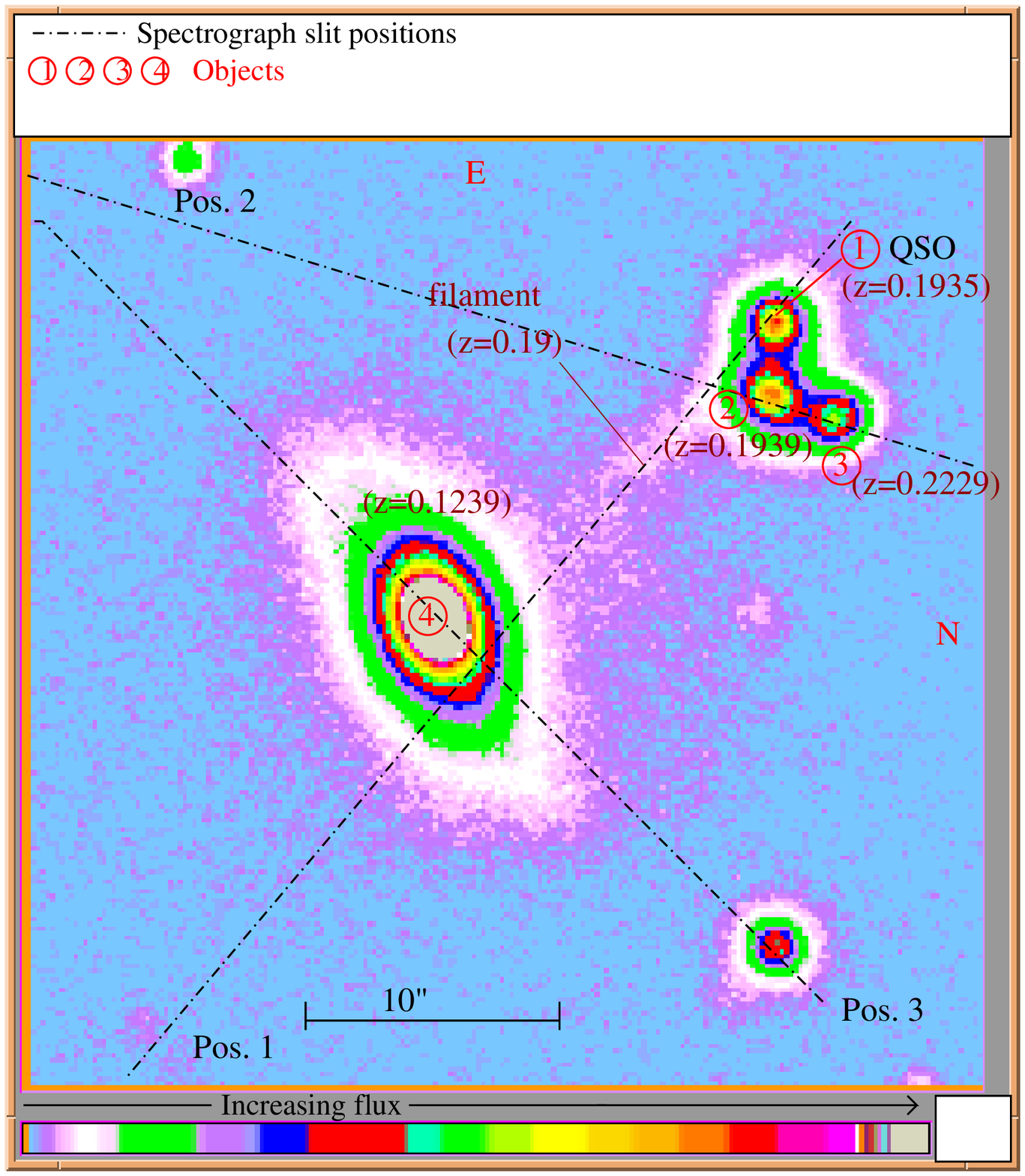}}
\hspace{1cm}\resizebox*{8.2cm}{6cm}{\includegraphics{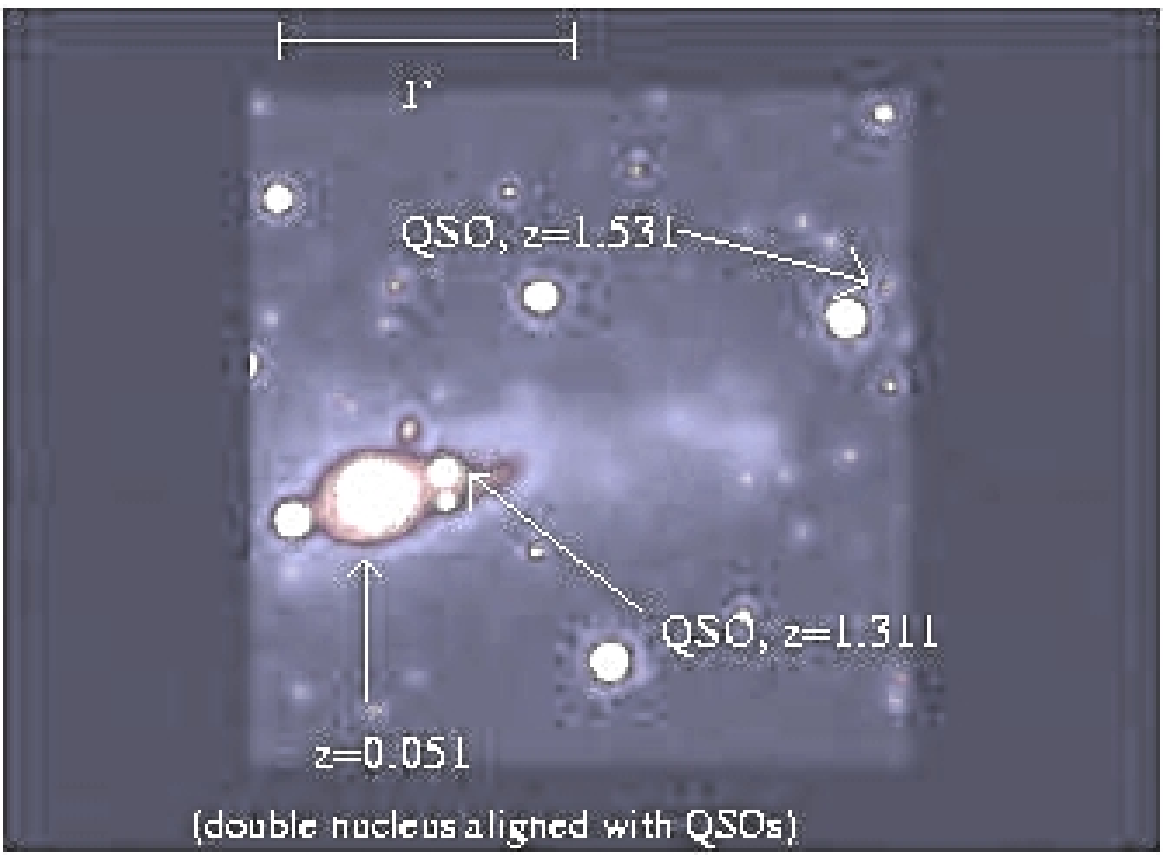}}\par}
\caption{Left: NEQ3; Sloan r' band, taken on the 2.6 m NOT (La Palma, Spain); 
reproduction of Fig. 1 of Guti\'errez \& L\'opez-Corredoira (2004). Right: 
GC 0248+430, a galaxy with two nuclei, and two QSOs, all of them aligned 
(+/- 5 degress); Sloan r' band, taken on the 2.6 m NOT; reproduction of 
Fig. 6 of L\'opez-Corredoira \& Guti\'errez (2006a).}
\label{Fig:angsize}
\end{figure*}       

The non-cosmological redshift hypothesis also affects galaxies differently 
from QSOs. Cases such as NGC 7603, AM 2004-295, AM 2052-221, NGC 1232, VV172, 
NEQ3, NGC 450/UGC 807, etc. present statistical anomalies also suggesting 
that the redshift of some galaxies different from QSOs might have 
non-cosmological redshifts. Not all supporters of the non-cosmological 
redshift agree with this idea; for instance, Arp claims that galaxies might 
have non-cosmological redshift because they derive from an evolution of 
ejected QSOs, while G. Burbidge only defends the non-cosmological redshifts 
in QSOs. In this paper, except for this paragraph, I shall talk only about 
anomalies in QSOs.

\section{Probabilities of being background QSOs}

There are two possible interpretations of these data: either QSOs with 
different redshifts are objects with different distances and the configurations 
are due to chance, or there are non-cosmological redshifts, and QSOs with 
different redshifts are at the same distance. The first position, the standard 
one, defends the hypothesis that in all cases the main galaxy is surrounded 
by background QSOs. The idea is quite straightforward. The position 
of anomalous redshifts is not naive enough to deny this possibility, and this 
might be the case in some examples. However, the question is not whether such 
a fortuitous projection is ``possible'' but whether it is ``probable''.

For the calculation of this probability $P$, it is normally assumed that the 
background/foreground objects in a small area are distributed according to a 
Poissonian distribution with the average density in any line of sight. There 
may be some clustering of QSOs, but this does not essentially affect the 
numbers. A conspiracy in which a given line of sight crosses several clusters 
of QSOs at different redshifts is not justified because the increase in 
probability due to the increase in density along lines of sight with clusters 
is compensated for by the additional factor to be multiplied by the present 
amount $P$ to take into account the probability of finding clusters in the line 
of sight. On average, in any arbitrary line of sight in the sky, the 
probability will be given anyway by the Poissonian calculation of $P$ with 
the average density of QSOs in the sky (see further details in subsection 
5.3.1 of L\'opez-Corredoira \& Guti\'errez 2004).

A much more important matter concerns the consideration of the number of 
events in the whole sky. Of course, there may be many objects that are quite 
peculiar but we must consider the global probability in the whole sky 
multiplying by the number of galaxies or QSOs as in the anomalous case. 
For instance, if we found an NGC galaxy of magnitude $m_g$ with a very low 
probability $P_0$ of being surrounded by $N$ QSOs up to some magnitude 
and angular distance, we must calculate the global probability $P$, 
multiplying it by the number of NGC galaxies (around eight thousand; or 
somewhat larger if we considered the southern hemisphere), or at least 
the galaxies in the whole sky up to magnitude $m_g$.

It is said that one should not carry out a calculation of the probability 
for a configuration of objects known a priori (for instance, that they form 
a certain geometrical figure) because, in some way, all possible 
configurations are peculiar and unique. That is right so long as we speak 
about random configurations that do not indicate anything special. For 
example, if the Orion constellation is observed and we want to calculate 
the chance of its stars being projected in that exact configuration, we 
will get a null probability (tending towards zero as the allowed error in the 
position of the stars with respect the given configuration goes to zero), 
but the calculation of this probability is worthless because we have selected 
a particular configuration observed a priori. Therefore, the statistics to be 
carried out should not be about the geometrical figure drawn by the sources, 
unless that geometrical configuration is representative of a physical process 
in an alternative theory (for instance, aligned sources might be 
representative of the ejection of sources by a parent source). 

In this last sense, I think that much of the statistics already published is 
valid and indicates the reality of some kind of statistical anomaly. 
It would be useful to look out for physical representations indicating 
peculiarities beyond mere uniqueness. I disagree with the claim that all 
attempts to calculate probabilities of unexpected anomalies are a posteriori 
and whose validity may therefore be rejected. Some astrophysicists, when 
looking at the images of the controversial objects, argue along the lines 
that the anomalous distributions of QSOs are curious, but that since they 
were observed their probability is 1 and there is therefore nothing special 
about them. According to this argument, everything is possible in a Poissonian 
distribution and nothing should surprise us. But I believe that statistics is 
something more serious than the postmodern rebuff that anything is possible. 

We think that this anti-statistical position, this way of rejecting the 
validity of the calculated probabilities, is equivalent to the scepticism 
that those unfamiliar with mathematics express when we discuss the low 
probability of winning the lottery. They continue to bet regardless with hope 
that, however low the probability, somebody is sure to win so why not me. 
Typically they are unaware of how low some probabilities are and make no 
distinction between a case such as $P\sim 10^{-2}$, 
which is a low but certainly makes a win possible from time to time, and the 
case $P\sim 10^{-7}$, which virtually ensures no wins during seven 
lifetimes of daily betting. Small numbers, like the huge numbers prevalent 
in astronomy, are not easily assimilated. Of course, somebody wins 
the lottery but this is because the number of players multiplied by the 
probability of winning of each player is a number not much lower than one; 
otherwise, nobody would ever be likely to win.

Even worse, imagine that a person wins the lottery four consecutive times 
with only one bet each time. If we did not believe in miracles, we might 
think that this person had cheated. We might carry out some statistical 
calculations and show how improbable it was that he/she could have won by 
chance. Somebody might say about these calculations that they are not valid 
because they were carried out a posteriori (after the person won the lottery 
four consecutive times). We would not agree because there is an alternative 
explanation (he/she is cheating; and this explanation could be thought of 
before the facts) and the event of winning the lottery four consecutive times, 
apart from being very peculiar among the random possibilities, would be an 
indication to support this hypothesis.

For our cases, we have facts (higher concentration of QSOs, alignments, 
QSOs projected onto filaments) which suggest that an alternative (a priori) 
theory claiming that galaxies/QSOs may be ejected by galaxies better 
represents the observations. The measured probabilities are not to form 
triangles or any shape observed a priori only because it was observed. 
The peculiarity that is analysed is not comparable with the previous example 
of Orion because we have in mind a physical representation rather than a 
given distribution of sources. The difference from the Orion problem is that 
the peculiarity of Orion is not associated with any peculiar physical 
representation to be explained by an alternative theory. The question is 
as follows: what is the probability, $P$, that the apparent fact be the fruit 
of a random projection of sources at different distances? In other words, what 
is the probability, $P$, that the standard theory can explain the observed 
facts without aiming at alternative scenarios?

There is some a posteriori information used normally in the calculations: 
for instance the maximum magnitude or distance of the QSOs according to 
what we have observed in our particular case. It is in this sense an a 
posteriori calculation, we are calculating the most pessimistic case 
(the lowest probability). Because of this, the values of $P$ might be 
slightly underestimated (by a factor not higher than 10-100) with respect 
to an a priori calculation without any information on magnitudes or radii, 
but values of $P$ lower than $\sim 10^{-4}$ should in any case be 
considered as statistically anomalous. In order to make a fairer estimate 
of the probability, we could calculate $P^*=2^n P$, where $n$ is 
the number of parameters on which $P$ depends. For instance, when we observe 
a source with magnitude 19 and we calculate $P(m<19)$ we are putting the 
limiting magnitude exactly at the observed number; a fairer calculation 
would be $P^*(m<(19+x))$ such that a source with magnitude 19 is a 
typical average source in the range $m<(19+x)$, i.e. 
roughly that half of the sources with $m<(19+x)$ have $m<19$ 
and the other half have $19<m<(19+x)$. This is equivalent to 
calculating $P^*(m<(19+x))=2 P(m<19)$ and 
for the correction we can multiply by a factor two for any independent 
parameter. Values of $P^*$ lower than around $10^{-3}$ should be 
considered as statistically anomalous.

\section{Gravitational lensing}

An explanation for anomalous redshift systems might be found in principle 
if we considered some kind of gravitational lensing by the foreground object. 
However, the effect on the enhancement of the probability produced by an 
individual galaxy is small. In order to increase by at least an order of 
magnitude in $P$ per object, i.e. an average enhancement of $\sim 10$ in 
density for each of the QSOs, we would need an average magnification of 
around 20,000 (L\'opez-Corredoira \& Guti\'errez 2004, sect. 5.3.2). 
This is so because the enhancement in the source counts increases because 
of the flux increase of each source but decreases owing to the area distortion, 
which reduces the number counts by losing the sources within a given area 
(Wu 1996). A magnification of $2\times 10^4$ is extremely high and impossible 
to achieve by a galaxy lens. The highest known values are up to a factor $\sim 
30$ (Ellis et al. 2001) for background objects apparently close to the central 
parts of massive clusters. Moreover, a single galaxy would only produce a 
significant magnification at very close distances (a few arcseconds) from 
the center. The possibility of multiple gravitational microlenses within the 
galaxy (Paczynski 1986) does not work either (Burbidge et al. 2005, sect. 5; 
L\'opez-Corredoira \& Guti\'errez 2006, sect. 8). 

Weak gravitational lensing by dark matter has also been proposed as the cause 
of the statistical correlations between low and high redshift objects, but 
this seems to be insufficient to explain them (Kovner 1989; Zhu et al. 1997; 
Burbidge et al. 1997; Burbidge 2001; Ben\'\i tez et al. 2001; Gazta\~naga 2003; 
Jain et al. 2003; Nollenberg \& Williams 2005; Tang \& Zhang 2005) and cannot 
work at all for the correlations with the brightest and nearest galaxies; 
L\'opez-Corredoira \& Guti\'errez (2007) have shown that gravitational 
lensing is not the solution for the possible minor axis excess of QSOs. 
Scranton et al. (2005) have claimed that the small amplitude correlation 
between QSOs and galaxies from the SDSS survey is due to weak gravitational 
lensing but this does not explain the most general case with bright nearby 
galaxies. Komberg \& Pilipenko (2008) suggested the existence of a large 
number of globular or proto-globular clusters in the intergalactic medium 
of clusters of galaxies as an explanation of the correlations, a hypothesis 
which is awaiting testing. In principle, it seems there are many field 
galaxies with an excess of surrounding QSOs. Further research is in any 
case necessary in some of these aspects.

\section{Are all QSOs with anomalies really QSOs?}

Even more important than thinking about gravitational lensing or discussing 
the probabilities of background projections is being sure that the 
identification of QSOs and their redshifts is correct. For instance, 
cases like 3C 343.1 (Arp et al. 2004a), if we believe that they are indeed 
a radio galaxy at z=0.34 and a radio QSO at z=0.75 separated by 0.25", 
are really spectacular, but are we sure of the correct identification 
of the sources? 

An example: Burbidge et al. (2003) suggested that many of the ultraluminous 
compact X-ray sources (ULXs) found in the main bodies of galaxies are ``local'' 
QSOs, or BL Lac objects, with high intrinsic redshifts in the process of 
being ejected from those galaxies. Certainly, there is an overdensity of 
these X-ray sources near galaxies but, before claiming a case of anomalous 
redshift, we have to be sure that they are indeed QSOs with different 
redshifts. Arp et al. (2004b) took some spectra of ULXs and saw that some 
of them are QSOs but others were not. L\'opez-Corredoira \& Guti\'errez 
(2006b) have shown that $>50$\% of ULXs are effectively QSOs but, except 
for few cases which are anomalous for other reasons (e.g., NGC 3628, 
NGC 4319), the probability of these QSOs being background objects is 
significant, while the cases with ULXs over the expected background 
were not QSOs. Therefore, there are not enough statistical anomalies to 
claim that some ULXs are non-cosmological redshift QSOs. 

Another example: Flesch \& Hardcastle (2004) published a catalog of 
candidate QSOs (with a probability $>40$\% of being QSOs) derived from the 
correlation of radio and X-ray sources with blue point-like optical objects. 
In this catalog, there is an overdensity of QSO candidates in fields near 
galaxies and for bright sources. However, L\'opez-Corredoira et al. (2008) 
showed that the probabilites of being QSOs were overestimated for bright 
objects and near galaxies. Therefore, again, there are in principle no 
reasons to think about statistical anomalies in this catalog.

\section{Discussion}

Some of the examples of apparent associations of QSOs and galaxies with 
different redshifts may be just fortuitous cases in which background 
objects are close to the main galaxy, although the statistical mean 
correlations remain to be explained, and some lone objects have a very low 
probability of being a projection of background objects. Nevertheless, 
these very low probabilities (down to $10^{-8}$ or even lower, assuming 
correct calculations) are not extremely low and, if the anomaly is real, one 
wonders why we do not find very clear anomalous cases with probabilities as 
low as $10^{-20}$. Gravitational lensing seems not to be a solution yet, 
although further research is required, and the aim that the probabilities
be calculated a posteriori is not in general an appropriate answer for 
avoiding or forgetting the problem. 

There are also other aspects of QSOs that are not well understood within the 
cosmological redshift assumption, and which could find an explanation within 
a non-cosmological redshift hypothesis (L\'opez-Corredoira \& Guti\'errez 
2006a, sect. 9): the extremely high luminosity of QSOs at high redshift 
and the absence of bright QSOs at low redshift, periodicity of redshifts, 
their age and metallicity and the lack of evolution signs, superluminal 
motions, spectral features in the emission and absorption lines that are 
not well understood, the mechanism of triggering activity, the fact that 
Faraday rotation does not increase with redshift, etc.

There are two possibilities: either all cases of associations are lucky 
coincidences with a higher probability than expected for some still unknown 
reason, or there are at least some few cases of non-cosmological redshifts. 
If we accepted that some objects (maybe not all of them) with different 
redshifts had the distance of the main galaxy, there might be some truth 
in those models (Burbidge 1999; Arp 1999a,b, 2001; Bell 2002a,b) in which 
QSOs and other types of galaxies are ejected by a parent galaxy, as proposed 
by Ambartsumian (1958). In these models, galaxies beget galaxies, not all the 
galaxies would be made from initial density fluctuations in a Big Bang 
Universe. For the explanation of the intrinsic redshift, there are several 
alternative hypotheses (reviews at Narlikar 1989; L\'opez-Corredoira 2003, 
sect. 2.1; 2006).

In my opinion, we must consider the question as an open problem to be solved. 
I maintain a neutral position, neither in favor of nor against 
non-cosmological redshifts. The debate has lasted a very long time, around 
40 years, and it would be time to consider making a last effort to finish 
with the problem. However, the scientific community does not seem very 
interested in solving the problem because most researchers consider it 
already solved. Supporters of the standard dogma of all redshifts being 
cosmological do not want to discuss the problem. Every time it is mentioned 
they just smile or talk about "a posteriori" calculations, manipulations of 
data, crackpot ideas, without even reading any paper on the theme. The 
Arp-Burbidge hypothesis has become a topic in which everybody has an 
opinion without having read the papers or knowing the details of the problem, 
because some leading cosmologists have said it is bogus. This means that it 
is very difficult to make any progress in this field, as is usual when a 
researcher is away from the mainstream (L\'opez-Corredoira \& Castro-Perelman, 
eds., 2008). On the other hand, the main supporters of the hypothesis 
of non-cosmological redshifts continue to produce tens of analyses of cases 
in favor of their ideas without too much care, pictures without rigorous 
statistical calculations in many cases, or with wrong identifications, 
underestimated probabilities, biases, use of incomplete surveys for 
statistics, etc., in many other cases. There are, however, many papers 
in which no objections are found in the arguments and they present quite 
controversial objects, but due to the bad reputation of the topic, the 
community simply ignores them. In this panorama, it would be difficult for 
the problem to be solved soon. Mainstream cosmologists are waiting for the 
death of the main leaders of the heterodox idea (mainly Arp and the couple 
Burbidge) to declare the idea as definitively dead. However, as in the case 
of Ambartsumian, some challenging ideas could survive or even be revived 
after some time if we leave open problems without a clear solution. 
Therefore, I would recommend that the community either finds good arguments 
against the Arp-Burbidge hypothesis, or that it allows their ideas to 
cohabit within the possible speculative hypotheses in cosmological 
scenarios.

\

{\bf Acknowledgements:}
Thanks are given to Terry Mahoney (IAC, Tenerife, Spain) for proof-reading 
of this paper, and Carlos M. Guti\'errez (IAC) for comments on its draft.

\end{document}